\documentclass{emulateapj}
\usepackage{times}
\usepackage{amssymb}

\newcommand{\calL}{{\cal L}}

\shorttitle{TINY HI CLOUDS}
\shortauthors{NAGASHIMA, INUTSUKA \& KOYAMA}

\begin{document}

\title{
How Long Can Tiny \ion{H}{1} Clouds Survive?}

\author{Masahiro Nagashima,\altaffilmark{1}
Shu-ichiro Inutsuka,\altaffilmark{1}
and 
Hiroshi Koyama\altaffilmark{2}}
\email{masa@scphys.kyoto-u.ac.jp}
\altaffiltext{1}{Department of Physics, Graduate School of Science,
Kyoto University, Sakyo-ku, Kyoto 606-8502, Japan}
\altaffiltext{2}{Department of Earth and Planetary Science, Kobe
University, Kobe 657-8501, Japan}

\begin{abstract}
We estimate the evaporation timescale for spherical \ion{H}{1} clouds
consisting of the cold neutral medium surrounded by the warm neutral
medium.  We focus on clouds smaller than 1pc, which corresponds to {\it
tiny \ion{H}{1} clouds} recently discovered by Braun \& Kanekar and
Stanimirovi{\'c} \& Heiles.  By performing one-dimensional spherically
symmetric numerical simulations of the two-phase interstellar medium
(ISM), we derive the timescales as a function of the cloud size and of
pressure of the ambient warm medium.  We find that the evaporation
timescale of the clouds of 0.01 pc is about 1Myr with standard ISM
pressure, $p/k_{B}\sim 10^{3.5}$ K cm$^{-3}$, and for clouds larger than
about 0.1 pc it depends strongly on the pressure.  In high pressure
cases, there exists a critical radius for clouds growing as a function
of pressure, but the minimum critical size is $\sim$ 0.03 pc for a
standard environment.  If tiny \ion{H}{1} clouds exist ubiquitously, our
analysis suggests two implications: tiny \ion{H}{1} clouds are formed
continuously with the timescale of 1Myr, or the ambient pressure around
the clouds is much higher than the standard ISM pressure.  We also find
that the results agree well with those obtained by assuming quasi-steady
state evolution.  The cloud-size dependence of the timescale is well
explained by an analytic approximate formula derived by Nagashima,
Koyama \& Inutsuka.  We also compare it with the evaporation rate given
by McKee \& Cowie.
\end{abstract}


\section{Introduction}
Recently discovered ``tiny \ion{H}{1} clouds'' have very small sizes,
$\sim 10^{-2}$ pc, and small \ion{H}{1} column density, $\sim 10^{18}$
cm$^{-2}$ \citep{bk05, sh05}.  This new population of clouds is
providing a challenge to our understanding of the evolutionary cycle of
the interstellar medium (ISM).  The origin of the clouds is still under
debate, but it is presumably the thermal instability \citep{f65, b86} in
turbulent gas \citep[e.g.][]{vpp95, ki02, ki04, kn02a, kn02b, ha05}
and/or the fragmentation of cold clouds crushed by interstellar shocks
\citep{nmkf05}.  Apart from the formation process, we can extract many
information about the surrounding ISM via investigation of the evolution
and statistics of the clouds.  In this {\it Letter}, we thus study the
evaporation rate and timescale of the tiny \ion{H}{1} clouds, which are
key quantities characterizing the fate of the clouds.

One of the simplest model is the clouds that consist of the cold neutral
medium (CNM) with temperature $T\sim 10^{1-2}$ K, surrounded by the warm
neutral medium (WNM) with $T\sim 10^{4}$ K, under pressure equilibrium.
The two phases are thermally stable balancing the heating rate due to,
e.g., photo-electric heating of dust grains, with the cooling rate due
to, e.g., line cooling by fine-structure transition of \ion{C}{2} atoms.
This is based on a widely accepted two-phase description of the ISM
\citep{fgh69, wolfire03}.  The evolution of clouds is, therefore,
described as the motion of the interface, or {\it front}, between the
CNM and WNM, driven by the thermal conduction.  Based on this picture,
\citet{zp69} and \citet{pb70} considered isobaric, steady motion of
fronts in plane-parallel geometry.  They clarified that the motion is
determined by pressure and that there is a {\it saturation} pressure for
which a static front can exist.  Their work has been substantially
extended by many authors \citep[e.g.][]{ers91, ers92, fs93, hp99}.

Compared to the analysis of the frontal motion in plane-parallel
geometry, the evolution of spherical clouds has been much less analyzed.
\citet{gl73} extended the work by \citet{zp69} and \citet{pb70} to
numerically compute isobaric flows in three dimensional spherical
geometry.  \citet{cm77} and \citet[][hereafter MC77]{mc77} considered a
model of spherical clouds surrounded by the hot ionized medium and
derived an analytic formula of the evaporation rate of clouds.  MC77
also considered cold clouds surrounded by the WNM, which can be directly
compared with the work by \citet{gl73}.  In our previous paper,
\citet{nki05}, we considered the growth of spherical clouds and showed a
new way to obtaining approximate analytic formula of the evaporation and
condensation rate.  However, because we adopted a simple polynomial form
of the cooling function proposed by \citet{ers92} for simplicity, our
understandings were only qualitative.  In this {\it Letter}, we shall
show the evaporation rate for a realistic cooling function by using a
full numerical simulation and compare it with the approximate analytic
formula and with the result of MC77.  We would like to stress that our
analysis can be applied to any spherical cold clouds surrounded by warm
gas under pressure equilibrium irrespective of their origin.  Thus the
results shown here are quite general.

Our aim is thus to estimate the evaporation timescale of cold \ion{H}{1}
clouds as a function of the cloud size and the pressure of the ambient
WNM.  This paper is outlined as follows.  In \S 2 we compute the
evaporation rate of cold clouds by using a full numerical simulation and
compare it with MC77's.  In \S3 we show the evaporation timescale.  In
\S4 we discuss the dependence of the timescale on the size and pressure.
We derive a new description of the evaporation rate as an improvement
upon MC77.  \S5 we provide conclusions.

\section{Evaporation rates}
Below we assume that clouds are spherically symmetric for simplicity, so
that we compute only radial structure of clouds.  Thus, the basic fluid
equations are written as
\begin{eqnarray}
 &&\frac{\partial\rho}{\partial t}+\frac{1}{r^2}\frac{\partial}{\partial r}r^2\rho v=0,\\
 &&\frac{\partial v}{\partial t}+v\frac{\partial v}{\partial r}=-\frac{1}{\rho}\frac{\partial p}{\partial r},\\
 &&\frac{\partial\rho e}{\partial t}+\frac{1}{r^2}\frac{\partial r^2\rho e v}{\partial r}+\frac{p}{r^2}\frac{\partial r^2 v}{\partial r}=-\rho\calL+\frac{1}{r^2}\frac{\partial}{\partial r}r^2\kappa(T)\frac{\partial T}{\partial r},\label{eqn:energy}
\end{eqnarray}
and the equation of state, $p={\rho}k_{B}T/{\mu}$, where
$\kappa(T)=2.5\times 10^3 \sqrt{T}$ erg$^{-1}$ cm$^{-1}$ K$^{-1}$
s$^{-1}$ is the conductivity for neutral gas \citep{p53}, $k_{B}$ the
Boltzmann constant, $\mu$ the mean molecular mass so that $\rho=\mu n$,
and $\calL$ is the heat-loss function defined as $\rho\calL\equiv
n^2\Lambda-n\Gamma$, where $\Lambda$ and $\Gamma$ are the cooling and
heating rates, respectively.  We adopt a simple analytic fitting
function given by \citet{ki02},
\begin{equation}
 \frac{\Lambda}{\Gamma}=10^7\exp\left(-\frac{118400}{T+1000}\right)+1.4\times 10^{-2}\sqrt{T}\exp\left(-\frac{92}{T}\right),
\end{equation}
and $\Gamma=2\times 10^{-26}$ erg s$^{-1}$, which are taking into
account many processes, e.g., photoelectric heating from dust grains and
PAHs, heating by cosmic rays and X-rays, and atomic line cooling from
hydrogen Ly$\alpha$ and \ion{C}{2} \citep{ki00}.  Throughout this paper,
we ignore the effect of magnetic fields.  This might alter the results
because the conductive heat flux is restricted along magnetic field
lines, and therefore it should be taken into account in future analysis.

Before performing full numerical simulations, we show the results for a
quasi-steady state (QSS), under which we transform the time derivative
to the spatial one, $\partial/\partial t\to
-\dot{R}_{c}\partial/\partial r$, where $R_{c}$ is the cloud size and
$\dot{R}_{c}$ is the velocity of the cloud size changing.  Then the
above equations become a set of ordinary differential equations.  The
details of the method to compute the structure will be found in a
separate paper \citep{nki06}.  In Fig.1, several snapshots of
evaporation rates, $\dot{M}(r)=4\pi r^2\rho(r) v(r)$, are shown by thin
dashed lines for the case of the saturation pressure, $p=p_{\rm
sat}\simeq 2823k_{B}$.  In contrast to the case of plane-parallel
geometry, the cloud evaporates by the heat conduction and the size
decreases as time passes.  As we shall show in \S\ref{sec:dis}, the
motion of the front shown in this figure is purely driven by the
curvature effect \citep{nki05}.

Next, we show the results obtained by full-numerical simulations. The
code is based on the second-order Godunov method \citep{vL77} in the
Lagrange coordinate.  We impose a constant pressure at the outer
boundary, that is, the WNM pressure is given as the boundary condition.
The initial structure of the cloud is computed by the QSS model with the
size of about 0.1 pc.  Correctly giving the initial condition is very
important for this computation.  If the deviation of the initial
condition from the correct structure is large, spherical sound waves
emerge and we could not extract the information on the evaporation.
Several snapshots at which the size of the cloud becomes the same as
that shown for the QSS are indicated by the thin solid lines in Fig.1.
It is evident that the results given by the full simulation agree well
with those given by assuming the QSS probably because the Mach number of
the flow is much below unity.  Therefore we can consider that the QSS is
a good assumption for the description of the evaporation of clouds.

The thick solid curve in Fig.1 denotes the evaporation rate as a
function of the cloud size, $\dot{M}(R_{c})$, where the edge of the
cloud is defined as a radius at which the density becomes $(n_{\rm
CNM}+n_{\rm WNM})/2$.  We have confirmed that the rate is proportional
to the size, $\dot{M}(R_{c})\propto R_{c}$.  In fact, this is the same
scaling against the size as that given by MC77.  Their evaporation rate
is
\begin{equation}
 \dot{M}_{\rm MC77}=\frac{16\pi\mu\kappa(T_{f}) R_{c}}{5k_{B}}=
  1.3\times 10^{15}T_{f}^{1/2}\left(\frac{R_{c}}{\rm pc}\right) {\rm ~g ~s}^{-1},
\end{equation}
where $T_{f}$ is the temperature of the WNM, $T_{f}\simeq 6.4\times
10^3$ K for the adopted heat-loss function.  This rate is plotted by the
thick dashed curve.  When we write our evaporation rate in a similar
way,
\begin{equation}
 \dot{M}=3.1\times 10^{14}T_{f}^{1/2}\left(\frac{R_{c}}{\rm pc}\right) {\rm ~g ~s}^{-1},
\end{equation}
therefore this is about a factor of four smaller than MC77's in spite of
the same dependence on the size.  The form of the evaporation rate shall
be discussed in \S\ref{sec:dis} in a different way from MC77.  Note that
the above rate is valid only for the case of the saturation pressure.
In the next section, we show the evaporation timescale for other cases.

\begin{figure}
\plotone{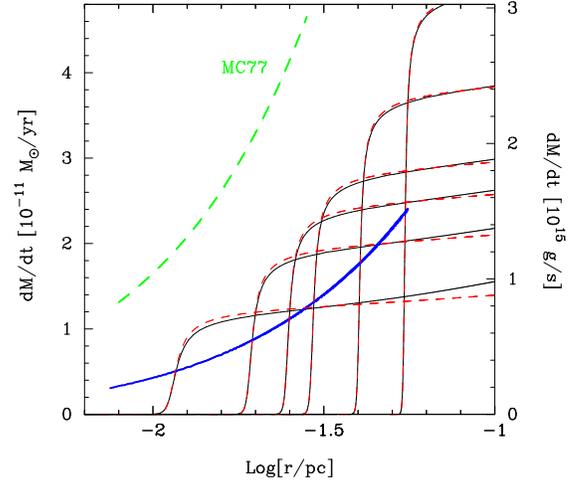}

\caption{Evaporation rate, $\dot{M}$ for the case of the saturation
pressure.  Thin solid and dashed lines represent snapshots of
$\dot{M}=4\pi r^{2}\rho(r)v(r)$ as a function of the distance from the
cloud center for the full numerical simulation and the QSS calculation,
respectively.  Thick solid curve indicates $\dot{M}$ as a function of
 the cloud size.  The size of clouds are defined in the text.  Thick
 dashed curve indicates $\dot{M}$ given by MC77.}

\label{fig:mdot}
\end{figure}

\section{Evaporation timescale}

Hereafter we define the evaporation timescale as $t_{\rm evap}\equiv
M/\dot{M}$, where the cloud mass $M$ is estimated by integrating the
mass density within the cloud radius, $R_{c}$, therefore $M_{c}\simeq
4\pi\rho_{\rm CNM}R_{c}^{3}/3$.  Fig.2 shows the evaporation timescale
against the cloud size $R_{c}$.  The solid straight line indicates
$t_{\rm evap}$ for the case of the saturation pressure, $p=p_{\rm sat}$,
and for the QSS.  The crosses on the solid line denote the result from
the full numerical simulation.  Similar to Fig.1, both are in excellent
agreement.  It is easily confirmed that $t_{\rm evap}\propto
R_{c}^{2}\propto M^{2/3}$ for $p=p_{\rm sat}$, reflecting the previous
result $\dot{M}\propto R_{c}$ and $M\propto R_{c}^{3}$.  Here it should
be worth noting that tiny \ion{H}{1} clouds with the size of $\sim
10^{-2}$ pc must disappear in $\sim$ 1Myr irrespective of the ambient
pressure.

The dashed and dot-dashed lines indicate the evaporation timescale for
the cases of $p/k_{B}=2000$ and 4000 K cm$^{-3}$, respectively.  Crosses
on the lines also denote the results from the full numerical
simulations.  Again we can see the good agreement.  Note that clouds
always evaporate for $p\leq p_{\rm sat}$, but large clouds can grow for
$p>p_{\rm sat}$, which is shown by the thin dot-dashed line at $R\ga
0.1$ pc.  This means that there exists a critical radius for growth
(condensation), $R_{\rm crit}$ (see the next section).  Hereafter we use
$t_{\rm evap}$ also as the growth timescale.  For these pressures, the
size dependence of $t_{\rm evap}$ is partly different from that for the
saturation pressure.  While it is very similar to that for the
saturation pressure, $t_{\rm evap}\propto R_{c}^{2}$ at $R_{c}\ll R_{\rm
crit}$, it becomes $t_{\rm evap}\propto R_{c}$ at $R_{c}\gg R_{\rm
crit}$.  This is very different feature from the expectation from MC77.
In the next section, we explain what the size dependence comes from.

\begin{figure}
\plotone{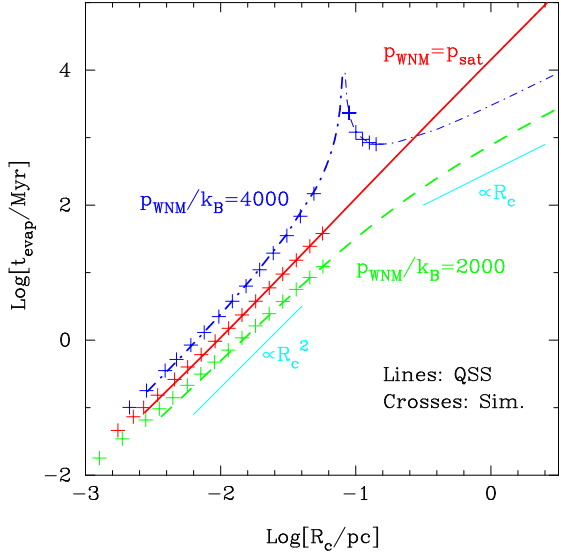}

\caption{Evaporation timescales defined as $M_{c}/\dot{M}$ as a function
of the radius of clouds.  Solid, dashed, and dot-dashed lines indicate
the timescale for $p=p_{\rm sat}$, 2000$k_{\rm B}$, and 4000$k_{\rm B}$,
respectively, under the assumption of the QSS.  Crosses denote
simulation results.  Note that $R_{c}\ga 0.1$ for $p_{\rm
WNM}/k_{B}=4000$ corresponds to the condensation timescale.}

\label{fig:tevap}
\end{figure}

\begin{figure}
\plotone{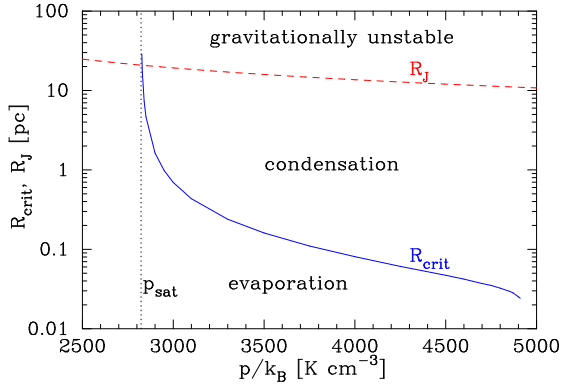}

\caption{Critical radius $R_{\rm crit}$ plotted by the solid line
against pressure of the WNM.  Jeans length $R_{\rm J}$ is also plotted
by the dashed line.}

\label{fig:Rcrit}
\end{figure}

\section{Discussion}\label{sec:dis}

Below we derive an approximate evaporation rate in the same way as that
of \citet{nki05}.  Firstly, we assume that the evolution is isobaric
because the motion of fluids is much slower than the sound speed.  Then the
energy equation (\ref{eqn:energy}) can be described as an evolution
equation of enthalpy,
\begin{equation}
\frac{\gamma}{\gamma-1}\frac{k_{B}}{\mu}\rho\left[\frac{\partial}{\partial t}+v\frac{\partial}{\partial r}\right]T
=-\rho\calL+\frac{1}{r^2}\frac{\partial}{\partial r}r^2\kappa\frac{\partial T}{\partial r}.\label{eqn:isobaric}
\end{equation}
Next, again because of the slow motion of fluids, we assume that the
fluids are in a QSS, so that the time derivative $\partial/\partial t$
can be replaced with $-\dot{R}_{c}\partial/\partial r$,
\begin{equation}
\frac{\gamma}{\gamma-1}\frac{k_{B}}{\mu}\rho u\frac{\partial T}{\partial r}
=-\rho\calL+\frac{\partial}{\partial r}\kappa\frac{\partial T}{\partial r}
+\frac{2}{r}\kappa\frac{\partial T}{\partial r},\label{eqn:qss}
\end{equation}
where $u\equiv v-\dot{R}_{c}$.  The first and second terms in the right
hand side emerge independent of geometry of fronts.  Therefore, if it
is reasonable to assume that the structure of fronts is almost
independent of geometry, we can replace the two terms with the left
hand side for the case of the plane-parallel geometry,
\begin{equation}
\frac{\gamma}{\gamma-1}\frac{k_{B}}{\mu}\rho u\frac{\partial T}{\partial r}
\simeq\frac{\gamma}{\gamma-1}\frac{k_{B}}{\mu}\rho u_{1}(p)\frac{\partial T}{\partial r}
+\frac{2}{r}\kappa\frac{\partial T}{\partial r},\label{eqn:qss2}
\end{equation}
where $u_{1}(p)$ is the fluid velocity with respect to the front rest
frame, and depends only on pressure.  Noting the fact that the
derivative of temperature, $\partial T/\partial r$ has a non-zero,
finite value only at the front, we can omit the derivative and
substitute $R_{c}$ for $r$,
\begin{equation}
u(R)\simeq u_{1}(p)
+\frac{\gamma-1}{\gamma}\frac{\mu}{k_{B}}\frac{2}{R_{c}}\frac{\kappa_{R}}{\rho_{R}},\label{eqn:front}
\end{equation}
where the subscript $R$ stands for values at the front, $r=R_{c}$.

To obtain a formula for the size evolution, we need a further assumption
that a convergence factor from $u$ to $\dot{R_{c}}$ is independent of
geometry, $f\equiv -\dot{R}_{c}/u$.  This becomes exact for the case of
the thin-front limit.  For three-dimensional spherical geometry, from
the mass-flux conservation, we obtain $-4\pi r^2\rho_{\rm
CNM}\dot{R}_{c}=4\pi R_{c}^2\rho_{R}u_{R}$, where $v=0$ inside the
cloud.  Then $f_{\rm sphere}=(R_{c}^2\rho_{R})/(r^2\rho_{\rm CNM})$.
For the plane-parallel geometry, $-\rho_{\rm CNM}c=\rho_{R}u_{R}$, where
$c$ is the speed of the front with respect to the cloud (CNM) rest
frame, which corresponds to $\dot{R}_{c}$.  Then $f_{\rm
p.p.}=\rho_{R}/\rho_{\rm CNM}$.  If the front is sufficiently thin, and
therefore the density $\rho$ is $\rho_{\rm CNM}$ even just inside the
front, we obtain $f_{\rm sphere}\to \rho_{R}/\rho_{\rm CNM}$.  This is
equal to $f_{\rm p.p.}$.  In reality, however, the front has a non-zero
thickness.  So the use of the same $f$ is an approximation.

Finally, we thus obtain the size evolution,
\begin{equation}
\dot{R}_{c}\simeq c(p)
-f\frac{\gamma-1}{\gamma}\frac{\mu}{k_{B}}\frac{2}{R_{c}}\frac{\kappa_{R}}{\rho_{R}},\label{eqn:size}
\end{equation}
where we explicitly write the argument of $c$, $p$, to stress that it
depends only on pressure.  For $p>p_{\rm sat}$, $c(p)>0$, and vice
versa.  This equation tells us the existence of a critical radius at
which $\dot{R}_{c}=0$ when $c(p)>0$,
\begin{equation}
 R_{\rm crit}\simeq f\frac{\gamma-1}{\gamma}\frac{\mu}{k_{B}}\frac{2}{c(p)}\frac{\kappa_{R}}{\rho_{R}}.\label{eqn:crit}
\end{equation}

The evaporation rate is obtained from the above equation,
\begin{equation}
\dot{M}\simeq -4\pi R_{c}^{2}\rho_{\rm CNM}\dot{R}_{c}
\equiv \dot{M}_{\rm p}+\dot{M}_{c},
\end{equation}
where $\dot{M}_{\rm p}$ and $\dot{M}_{\rm c}$ are the ``pressure'' and
``curvature'' terms, respectively, and
\begin{eqnarray}
 \dot{M}_{\rm p}&=&-4\pi R_{c}^{2}\rho_{\rm CNM}c(p)\propto R_{c}^{2},\\
 \dot{M}_{\rm c}&=&4\pi R_{c}^{2}\rho_{\rm CNM}f\frac{\gamma-1}{\gamma}\frac{\mu}{k_{B}}\frac{2}{R_{c}}\frac{\kappa_{R}}{\rho_{R_{c}}}\propto R_{c}.\label{eqn:mdot}
\end{eqnarray}
From the different dependences on $R_{c}$, we can see that the pressure
term dominates for large clouds, $R_{c}\gg R_{\rm crit}$, and that the
curvature term does for small clouds, $R_{c}\ll R_{\rm crit}$.  Using
$f\simeq\rho_{R}/\rho_{\rm CNM}$ and $\gamma=5/3$, we obtain
\begin{eqnarray}
 \dot{M}_{\rm c}&=&\frac{16\pi\mu\kappa_{R}R_{c}}{5k_{B}}.\label{eqn:mdot2}
\end{eqnarray}
Thus the difference from MC77's result for the curvature term is
\begin{equation}
 \frac{\dot{M}_{c}}{\dot{M}_{\rm MC77}}=\frac{\kappa_{R}}{\kappa(T_{f})}=\sqrt{\frac{T_{R}}{T_{f}}},
\end{equation}
because of $\kappa\propto\sqrt{T}$.  Thus the difference between our
result and MC77's can be partly explained by the above in addition to
the thin-front approximation to derive the curvature term, $\dot{M}_{\rm
c}$.  MC77 took into account only the curvature term.

The size dependence of the evaporation timescale is thus dependent on
the size.  For small clouds, $R_{c}\ll R_{\rm crit}$, the timescale is
$t_{\rm evap}\sim M/\dot{M}_{\rm c}\propto R_{c}^{2}$.  For large
clouds, $R_{c}\gg R_{\rm crit}$, it is $t_{\rm evap}\sim M/\dot{M}_{\rm
p}\propto R_{c}$.  Thus the size dependence of the evaporation timescale
shown in Fig.\ref{fig:tevap} is well explained by the above approximate
evaporation rate we derived.

A further important point is the relationship between the critical
radius, $R_{\rm crit}$, and the Jeans length for gravitational
instability, $R_{\rm J}\equiv c_{S}\sqrt{\pi/G\rho}$, where $c_{S}$ is
the sound velocity.  Fig.\ref{fig:Rcrit} shows $R_{\rm crit}$ and
$R_{\rm J}$ as a function of the ambient pressure.  We find that there
can exist clouds growing only by condensation for the case of $p>p_{\rm
sat}$.  The timescale until the cloud size exceeds $R_{\rm J}$ is,
however, quite long, as shown in Fig.\ref{fig:tevap}.

\section{Conclusion}
We have investigated the evaporation rate and timescale of tiny
\ion{H}{1} clouds by using numerical simulations.  We confirmed that the
results are almost recovered by assuming the evolution under the
quasi-steady state because the fluid velocity is much slower than the
sound velocity.  We have found that clouds with a size of about 0.01 pc
evaporate in approximately 1Myr almost independent of pressure of the
ambient WNM.  This suggests that there might be some mechanisms to
continuously form tiny \ion{H}{1} clouds, or, that the ambient pressure
around the clouds is much higher than the standard ISM pressure, if
their existence is ubiquitous.  The evaporation timescale of clouds
larger than 0.1 pc, however, depends strongly on the pressure.  Clouds
larger than a critical radius can even grow if the pressure is higher
than the saturation pressure.

In order to physically understand the simulation results, we derived an
analytic formula for evaporation by assuming the isobaric evolution and
a structure of the interface independent of geometry.  The obtained
evaporation rate has two separate terms: ``pressure'' and ``curvature''
terms.  The former is independent of geometry, that is, it emerges even
in analysis for plane-parallel geometry, and it becomes zero for the
saturation pressure.  The latter is proportional to the size of clouds,
that is, the curvature.  We have found that the obtained evaporation
rate is a natural extension of an evaporation rate obtained by MC77,
which corresponds to our ``curvature'' term, but MC77's rate is a little
higher than ours.  For lower pressure than the saturation pressure, the
signs of the two term are the same.  Therefore clouds always evaporate.
For higher pressure, on the other hand, the signs are different.
Therefore a critical size exists and the size of clouds determines
whether they grow or evaporate.  We have confirmed that the simulation
results show the same dependence on the size as the analytic formula.

We would like to stress that our analysis can be adapted irrespective of
the origin of the tiny \ion{H}{1} clouds.  In order to obtain
information on the formation process, statistical properties such as
mass function of clouds should be required.  This will be done in
future.

\section*{ACKNOWLEDGMENTS}    

This work was supported by the Grant-in-Aid for the 21st Century COE
"Center for Diversity and Universality in Physics" from the Ministry of
Education, Culture, Sports, Science and Technology (MEXT) of Japan, and
by the Astronomical Data Analysis Center (ADAC) of the National
Astronomical Observatory, Japan.  Numerical simulation was in part
carried out on the general common-use computer system at the ADAC.  MN
is supported by the Japan Society for the Promotion of Science for Young
Scientists (No.207).


\begin{thebibliography}{}   
\bibitem
	[Balbus(1986)]{b86}
	Balbus, S. A. 1986, \apjl, 303, L79
\bibitem[Braun \& Kanekar(2005)]{bk05}
	Braun, R., \& Kanekar, N. 2005, \aap, 436, L53
\bibitem[Cowie \& McKee(1977)]{cm77}
	Cowie, L. L., \& McKee, C. F. 1977, \apj, 211, 135
\bibitem
	[Elphick, Regev \& Spiegel(1991)]{ers91}
	Elphick, C., Regev, O., \& Spiegel, E. A. 1991, \mnras, 250, 617
\bibitem
	[Elphick, Regev \& Shaviv(1992)]{ers92}
	Elphick, C., Regev, O., \& Shaviv, N. 1992, \apj, 392, 106
\bibitem
	[Field(1965)]{f65}
	Field, G. B. 1965, \apj, 142, 531
\bibitem[Ferrara \& Shchekinov(1993)]{fs93}
	Ferrara, A., \& Shchekinov, Yu. 1993, \apj, 417, 595
\bibitem[Field, Goldsmith \& Habing(1969)]{fgh69}
	Field, G. B., Goldsmith, D. W., \& Habing, H. J. 1969, \apjl, 155, L149
\bibitem
	[Graham \& Langer(1973)]{gl73}
	Graham, R., \& Langer, W. D. 1973, \apj, 179, 469
\bibitem[Hennebelle \& Audit(2005)]{ha05}
	Hennebelle, P., \& Audit, E. 2005, \aap, 433, 1
\bibitem[Hennebelle \& P{\'e}rault(1999)]{hp99}
	Hennebelle, P., \& P{\'e}rault, M. 1999, \aap, 351, 309
\bibitem
	[Koyama \& Inutsuka(2000)]{ki00}
	Koyama, H., \& Inutsuka, S. 2000, \apjl, 532, 980
\bibitem
	[Koyama \& Inutsuka(2002)]{ki02}
	Koyama, H., \& Inutsuka, S. 2002, \apjl, 564, L97
\bibitem
	[Koyama \& Inutsuka(2004)]{ki04}
	Koyama, H., \& Inutsuka, S. 2004, \apjl, 602, L25
\bibitem
	[Kritsuk \& Norman(2002a)]{kn02a}
	Kritsuk, A. G., \& Norman, M. L. 2002a, \apjl, 569, L127
\bibitem
	[Kritsuk \& Norman(2002b)]{kn02b}
	Kritsuk, A. G., \& Norman, M. L. 2002b, \apjl, 580, L51
\bibitem
	[McKee \& Cowie(1977)]{mc77}
	McKee, C. F., \& Cowie, L. L. 1977, \apj, 215, 213
\bibitem[Nagashima, Koyama \& Inutsuka(2005)]{nki05}
	Nagashima, M., Koyama, H., \& Inutsuka, S. 2005, \mnras, 361, L25
\bibitem[Nagashima, Koyama \& Inutsuka(2006)]{nki06}
	Nagashima, M., Koyama, H., \& Inutsuka, S. in preparation
\bibitem[Nakamura et al.(2005)]{nmkf05}
	Nakamura, F., McKee, C. F., Klein, R. I., \& Fisher, R. T. 2005,
	preprint (astro-ph/0511016)
\bibitem[Parker(1953)]{p53}Parker, E.N. 1953, \apj, 117, 431
\bibitem
	[Penston \& Brown(1970)]{pb70}
	Penston, M. V., \& Brown, F. E. 1970, \mnras, 150, 373
\bibitem[Stanimirovi{\'c} \& Heiles(2005)]{sh05}
	Stanimirovi{\'c}, S., \& Heiles, C. 2005, \apjl, 631, L371
\bibitem[van Leer(1977)]{vL77}
	van Leer, B. 1977, J. Comp. Phys., 135, 229
\bibitem[V{\'a}zquez-Semadeni, Passot \& Pouquet(1995)]{vpp95}
	V{\'a}zquez-Semadeni, E., Passot, T., \& Pouquet, A. 1995, \apj,
	441, 702
\bibitem[Wolfire et al.(2003)]{wolfire03}
	Wolfire, M. G., McKee, C. F., Hollenbach, D., \& Tielens, A. G. G. M.
	2003, \apj, 587, 278
\bibitem
	[Zel'dovich \& Pikel'ner(1969)]{zp69}
	Zel'dovich, Ya. B., \& Pikel'ner, S. B. 1969, JETP, 29, 170
\end{thebibliography}
\end{document}